\documentclass[preprint,aps]{revtex4}

\usepackage{amsmath}
\usepackage{amssymb}
\usepackage{graphicx}% Include figure files
\usepackage{dcolumn}% Align table columns on decimal point
\usepackage{bm}% bold math
\begin{document}
%\preprint{Submitted to }
%\thispagestyle{empty}
%%%%%%%%%%%%%%%%%%%%%%
%\draft
%\preprint{submitted to ...}
%%%%%%%%%%%%%%%%%%%%%%%%%%%%%%%%%%%%%%%%%%%%%%%%%%%%%%%%%%%%%%%%%%%%%%
% TITLE  %%%%%%%%%%%%%%%%%%%%%%%%%%%%%%%%%%%%%%%%%%%%%%%%%%%%%%%%%%%%%
%%%%%%%%%%%%%%%%%%%%%%%%%%%%%%%%%%%%%%%%%%%%%%%%%%%%%%%%%%%%%%%%%%%%%%
%
\title{Reaction under vacancy-assisted diffusion at high quencher concentration} %mediated reaction 
\author{Kazuhiko Seki}
\email{k-seki@aist.go.jp}
\affiliation{
National Institute of Advanced Industrial Science and Technology (AIST)\\
AIST Tsukuba Central 5, Higashi 1-1-1, Tsukuba, Ibaraki, Japan, 305-8565
}
\author{M. Tachiya
}
\email{m.tachiya@aist.go.jp}
\affiliation{
National Institute of Advanced Industrial Science and Technology (AIST)\\
AIST Tsukuba Central 5, Higashi 1-1-1, Tsukuba, Ibaraki, Japan, 305-8565
}

%\date{\today}
%%%%%%%%%%%%%%%%%%%%%%%%%%%%%%%%%%%%%%%%%%%%%%%%%%%%%%%%%%%%%%%%%%%%%%
% ABSTRACT  %%%%%%%%%%%%%%%%%%%%%%%%%%%%%%%%%%%%%%%%%%%%%%%%%%%%%%%%%%
%%%%%%%%%%%%%%%%%%%%%%%%%%%%%%%%%%%%%%%%%%%%%%%%%%%%%%%%%%%%%%%%%%%%%%
%\thispagestyle{empty}
\begin{abstract}

Theory of diffusion-mediated reactions is already established for 
the target problem in the dilute limit, 
where the immobile target is surrounded by many quenchers. 
For lattice random walks in the crowded situation, each quencher is surrounded by other quenchers differently. 
As a result, each quencher  
migrates differently in the presence of site blocking effects. 
However, in the conventional theory, such difference is ignored 
and quenchers are assumed to move independently of each other. 
In this paper, 
theory of diffusion-mediated reactions of target problem is developed 
by taking into account the site blocking effects 
for quencher migration and 
the difference in the configuration of quenchers around each quencher.  
Our result interpolates between those in high and low limits of quencher concentrations 
and is a lower bound of the survival probability. 
In the static limit, the exact result is reproduced for a localized sink. 
In the presence of diffusion, 
the approximation is better when intrinsic reaction rates are low.  
%\begin{center}
%Abstract
%\end{center}
\end{abstract}
%\PACS 78.55.Qr \sep 72.20.Ee \sep  05.40.a

\maketitle
\newpage
%%%%%%%%%%%%%%%%%%%%%%%%%%%%%%%%%%%%%%%%%%%%%%%%%%%%%%%%%%%%%%%%%%%%%%
% Introduction %%%%%%%%%%%%%%%%%%%%%%%%%%%%%%%%%%%%%%%%%%%%%%%%%%%%%%%%%%%%%%
%%%%%%%%%%%%%%%%%%%%%%%%%%%%%%%%%%%%%%%%%%%%%%%%%%%%%%%%%%%%%%%%%%%%%%
\setcounter{equation}{0}
\section{Introduction}
\vspace{0.5cm}

Theory of diffusion-mediated reactions is already established for 
the target problem in the dilute limit. 
\cite{Rice,Inokuti,Tachiya74,Tachiya83,Szabo89}
%%% revised 
Consider reaction between excited probe $\mbox{A}^*$ and 
quencher $\mbox{B}$ which deactivates excited probe. 
In ordinary experimental conditions, 
$\mbox{A}^*$ is minority species and $\mbox{B}$ is majority species. 
The case in which minority species are immobile and majority species are mobile is in general called 
the target problem. 
The opposite case is called the trapping problem. 
In this paper, we consider the target problem. 
We are interested in the decay of fraction of $\mbox{A}^*$ 
after pulsed excitation or the fraction of 
$\mbox{A}^*$ when $\mbox{A}$ is continuously excited. 
%%% revised 
In the dilute limit, 
movement of each quencher can be regarded as an independent event. 
When the target is surrounded by $N$ quenchers on the lattice with $M$ sites 
the survival probability of the target at time $t$ denoted by 
$P_N (t)$ is expressed in terms of 
the pair survival probability at time $t$ 
of a quencher starting from $\vec{r}_\ell$, $f (\vec{r}_\ell, t)$, by \cite{Tachiya83}
\begin{eqnarray}
P_N (t) &=&  \left( \sum_{\ell=1}^M  f (\vec{r}_\ell, t)  \right)^N \nonumber \\
&=& \left( 1 - \frac{1}{M} 
\sum_{\ell=1}^M  \left[ 1- f (\vec{r}_\ell, t) \right]  
\right)^N 
\nonumber \\
&\simeq& \exp \left( -c \sum_{\ell=1}^\infty \left[ 1- f (\vec{r}_\ell, t) \right] \right), 
\label{intro}
\end{eqnarray}
where the infinite limit of all lattice points, $M$, is taken 
and the concentration is given by, $c=\lim_{M \rightarrow \infty} N/M$ 
in the thermodynamic limit. 
The theory is applicable even under 
the long-range reactions and the presence of electrostatic potential 
among reactants, 
which can be taken into account in 
$f (\vec{r}_\ell, t)$ as long as quencher concentration is dilute. 

In the conventional theory, 
the decay of the survival probability by bulk reactions 
has been formulated in terms of the pair probability of  
the target and a quencher by ignoring the excluded volume interactions among quenchers. 
However, in the crowded situation, each quencher is surrounded by other quenchers differently. 
As a result, each quencher  
migrates differently in the presence of site blocking effects. 
In the conventional theory, such difference is ignored and 
quenchers are assumed to move independently of each other. 
In this paper, we take into account the site blocking effects for quencher migration and 
the difference in the configuration of quenchers around each quencher by
applying Nakazato-Kitahara's theory of tracer diffusion. \cite{Nakazato}
%%% revised 
Excluded volume interactions are taken into account 
by prohibiting double occupancy of quenchers 
in site blocking effects. 
Quenchers can jump only to the empty neighbor sites. 
By noticing the success of Nakazato-Kitahara's interpolating 
formula on 
the tracer diffusion coefficient 
between low and high concentrations of 
diffusing particles, 
we apply it to the target reaction on a lattice.  
Though site blocking is an aspect of many body interactions, 
rigorous results can be obtained by this method. 
In the continuous space, 
excluded volume interactions can be taken into account by 
introducing short range repulsive potentials 
and applying decoupling approximation of 
density correlations, 
as shown by Kuzovkov {\it et al.} \cite{Kuzovkov}
%%% revised 

%\newpage
%%%%%%%%%%%%%%%%%%%%%%%%%%%%%%%%%%%%%%%%%%%%%%%%%%%%%%%%%%%%%%%%%%%%%%
% Static quenching at high quencher concentration %%%%%%%%%%%%%%%%%%%%%%%%%%%%%%%%%%%%%%%%%%%%%%%%%%%%%%%%%%%%%%
%%%%%%%%%%%%%%%%%%%%%%%%%%%%%%%%%%%%%%%%%%%%%%%%%%%%%%%%%%%%%%%%%%%%%%
%\setcounter{equation}{0}
\section{Static quenching without double occupancy}
\vspace{0.5cm}

The simplest results which prohibit double occupancy of the same site 
are obtained in the absence of diffusion. 
Although the results are known, we rederive them to illustrate our method. 

We consider the lattice with $M$ sites. 
There are $N$ quenchers on the lattice.
We indicate the configuration of quenchers  
by the set of vectors denoting the lattice sites occupied by quenchers,  
$(\vec{r}_1, \vec{r}_2, \cdots , \vec{r}_N)$. 
If quenchers are initially randomly distributed, 
the probability of taking an initial configuration $(\vec{r}_1, \vec{r}_2, \cdots , \vec{r}_N)$ 
is given by,  
\begin{eqnarray}
P(\vec{r}_1, \vec{r}_2, \cdots , \vec{r}_N; 0)
= 1/\left(_{M} C_N \right). 
\label{initialprob}
\end{eqnarray}
The survival probability is obtained 
by applying the Cauchy's integral theorem, \cite{Arfken} 
\begin{eqnarray}
P_N (t) = \frac{1}{2 \pi i} \int d\, x \frac{1}{x^{N+1}} \frac{1}{_{M} C_N } \prod_{\ell=1}^M \left(1+ x \exp (- k \left(\vec{r}_\ell \right) t )\right) ,  
\label{staticinv1}
\end{eqnarray} 
where the path of integration encircles the origin on the complex plane 
and 
the right hand side of Eq. (\ref{staticinv1}) represents 
the joint probability of independent quenching events at time $t$ from all 
possible quencher configurations. 
Eq. (\ref{staticinv1}) can be rewritten as, 
\begin{eqnarray}
P_N (t) = \frac{1}{2 \pi i} \int d\, x \frac{1}{x^{N+1}} \frac{1}{_{M} C_N } \exp
\left[\sum_{\ell=1}^M \ln  \left(1+ x \exp (- k \left(\vec{r}_\ell \right) t )\right) \right] .  
\label{staticinv2}
\end{eqnarray} 
In the thermodynamic limit, Eq. (\ref{staticinv2}) is simplified 
by introducing Stirling formula $n!=\sqrt{2 \pi n} \exp (-n) n^n$ and applying the steepest descent method, 
\begin{eqnarray}
P_N (t) &=& \frac{1}{2 \pi i} \int d\, x \frac{(1+x)^M}{x^{N+1}} \frac{1}{_{M} C_N } \exp
\left[\sum_{\ell=1}^M \ln  \left(1+ \frac{x}{1+x}
\left( \exp (- k \left(\vec{r}_\ell \right) t )-1 \right) \right) \right]\\
&=& \exp \left[ \sum_{\ell=1}^M  \ln\left\{ 1 +  c\left( \exp \left(- k \left(\vec{r}_\ell \right) t \right) -1 \right)\right\} \right] , 
\label{staticinv3}
\end{eqnarray} 
where $c=N/M$ is the concentration. 
Eq. (\ref{staticinv3}) is the known expression for the static quenching obtained by 
Allinger and Blumen (AB) using a different method. 
\cite{Allinger}
Our method is not simple but shows that 
Eq. (\ref{staticinv3}) is correct in the thermodynamic limit for any concentration at 
all times and thus confirms the conclusion derived from the AB method, 
where the occupancy probability at each lattice site is assumed to be $c$ instead of 
random occupancy of quencher sites among available lattice sites. 
In the AB method, 
the number of quenchers for a finite lattice with $M$ sites is not necessarily equal to $N$ 
since the occupancy probability of each lattice site is given for each realization of quencher configurations. 
For $c=1$ we find the familiar result of, 
\begin{equation}
P_N (t) = \exp \left( - \sum_{\ell=1}^M k \left(\vec{r}_\ell \right) t \right), 
\label{staticfamiliar}
\end{equation}
and its Laplace transform is given by, 
\begin{equation}
\hat{P}_N (s)  = 1/\left(s+  \sum_{\ell=1}^M k \left(\vec{r}_\ell \right) t \right). 
\label{staticfamiliarL}
\end{equation}
In the opposite limit, $c \sim 0$, 
Eq. (\ref{staticinv3}) reduces to the well-known result, \cite{Tachiya74,Allinger}
\begin{equation}
P_N (t) \simeq \exp  \left[ - \sum_{\ell=1}^M c\left(1- \exp \left(- k \left(\vec{r}_\ell \right) t \right)  \right) \right]  . 
\label{staticknown}
\end{equation}
Eq. (\ref{staticknown}) is the static limit of Eq. (\ref{intro}).

When reaction takes place only at a target site $\vec{r}_R$, 
$k \left( \vec{r}_\ell \right) = k_0 \delta_{\vec{r}_\ell, \vec{r}_R}$, 
the survival probability, Eq. (\ref{staticinv3}), is simplified to, 
\begin{eqnarray}
P_N (t) = 1-c + c \exp (- k_0 t )  ,  
\label{static0}
\end{eqnarray}
and the Laplace transform, 
$\hat{P}_N \left(  s \right) = \int_0^\infty \,dt \exp (-st) P_N \left(  t \right)$, 
is expressed as, 
\begin{eqnarray}
\hat{P}_N (s) = \frac{s+(1-c)k_0}{s\left(s+k_0 \right)} . 
\label{staticLaplace0}
\end{eqnarray}
These trivial results will be used to check the  
results obtained under the presence of correlated diffusion. 
In this model, 
if a target site is not occupied by a quencher, 
the reaction never occurs there. 
However, 
in the presence of diffusion, 
even if a target site is not initially occupied by a quencher, 
a quencher may come to it by diffusion and react with it. 
In other words, 
the survival probability obtained under the condition of static quenching is 
always higher than 
that in the presence of diffusion. 

Before closing this section, 
we comment on the natural decay. 
When the natural decay of the target with the time constant $\tau_0$ is present, we multiply 
$P_N (t)$ by 
$\exp (- t/\tau_0)$ and $s$ changes to $s+ 1/\tau_0$ in 
$\hat{P}_N (s)$. 
The natural decay of the quencher with the time constant $\tau_{0q}$ is taken into account by 
replacing $k \left(\vec{r}_\ell \right)$ by $k \left(\vec{r}_\ell \right)+ 1/\tau_{0q}$. 
The natural decay can be taken into account even under the presence of correlated diffusion 
in the same way. 

%\newpage
%%%%%%%%%%%%%%%%%%%%%%%%%%%%%%%%%%%%%%%%%%%%%%%%%%%%%%%%%%%%%%%%%%%%%%
% Qenching with hindered diffusion %%%%%%%%%%%%%%%%%%%%%%%%%%%%%%%%%%%%%%%%%%%%%%%%%%%%%%%%%%%%%%
%%%%%%%%%%%%%%%%%%%%%%%%%%%%%%%%%%%%%%%%%%%%%%%%%%%%%%%%%%%%%%%%%%%%%%
%\setcounter{equation}{0}
\section{Quenching under diffusion with excluded volume interactions}
\vspace{0.5cm}

Quenchers perform random walk on a lattice under the condition 
that each site cannot be occupied by more than one quencher at the same time. 
Quenchers can jump only to the empty neighbor sites. 
The movement of quencher is influenced by the position of other quenchers 
through the site blocking effects. 
As a result,   
quencher diffusion is highly correlated at high concentrations. 
In addition to correlated diffusion, 
reaction takes place depending on the distance between 
the quencher and the target. 
Since the target is immobile, 
the reaction rate depends only on the configuration of quenchers.

As before, we consider the lattice with $M$ quencher sites. 
An excited target is located at the origin. 
There are $N$ quenchers on the lattice.
If quenchers are initially randomly distributed over available sites, 
the probability of taking an initial configuration 
$(\vec{r}_1, \vec{r}_2, \cdots , \vec{r}_N)$ 
is given by 
Eq. (\ref{initialprob}). 
The survival probability is given by, 
\begin{eqnarray}
P_N (t) = \sum_{\{ \vec{r}_i\}} P(\vec{r}_1, \vec{r}_2, \cdots , \vec{r}_N; t), 
\label{survivaldef}
\end{eqnarray}
where the summation should be taken over all possible 
quencher configurations. 

The self-diffusion of correlated random walk is studied by Nakazato and Kitahara 
in the absence of reaction. \cite{Nakazato} 
Site blocking effects on the diffusion of tagged particle is calculated. \cite{Nakazato,Suzuki,Okamoto07}
Following them, we  introduce ket vectors for 
all accessible sites of quenchers. 
The ket vector $| \vec{r}, \bullet \rangle$ denotes the occupation 
of site $\vec{r}$ 
by a quencher particle, 
and $| \vec{r}, \phi \rangle$ represents that site $\vec{r}$ is an empty site. 
The probability of finding a configuration 
$\left( \vec{r}_1, \cdots \vec{r}_N \right)$, at time $t$ 
averaged over all possible initial configurations of random occupation 
is written as, 
\begin{eqnarray}
P(\vec{r}_1, \cdots , \vec{r}_N; t) &=& e^{-t/\tau_0} 
\left( \prod_{\ell=1}^N \langle \vec{r}_\ell, \bullet | \right) \left( \prod_{k=N+1}^M \langle \vec{r}_k, \phi | \right) e^{Ht} 
\sum_{\left\{i \right\}} 
\frac{1}{_M C_N}
\left( \prod_{\ell=1}^N |\vec{r}_\ell, \bullet \rangle \right) \left( \prod_{k=N+1}^M | \vec{r}_k, \phi \rangle \right), 
\nonumber \\
\end{eqnarray}
where the sum is taken over all possible configurations of $N$ occupied sites on the $M$ sites. 
$H$ is given by $H=H_{\rm w} + H_{\rm rc}$, 
where $H_{\rm w}$ describes the diffusion of quenchers, \cite{Nakazato,Suzuki,Okamoto07}
\begin{eqnarray}
H_{\rm w} = \sum_{\langle n,m \rangle} \Gamma/(2d)
\left( 
|\vec{r}_n, \bullet \rangle \langle \vec{r}_n, \phi | 
\cdot %\otimes 
|\vec{r}_m, \phi \rangle \langle \vec{r}_m, \bullet | 
- |\vec{r}_n, \bullet \rangle \langle \vec{r}_n, \bullet | 
\cdot %\otimes 
|\vec{r}_m, \phi \rangle \langle \vec{r}_m, \phi | 
\right), 
\end{eqnarray}
where $\Gamma$ is the jump frequency of quencher 
and the sum is taken over all nearest neighbor pairs of 
accessible lattice sites by   
quenchers. 
Transition  is possible from the state  
$| \vec{r}_m, \bullet \rangle | \vec{r}_n, \phi \rangle$ to the state 
$| \vec{r}_n, \bullet \rangle | \vec{r}_m, \phi \rangle$,  
which indicates that the site $n$ must be empty to accept a quencher from an occupied neighboring site. 
Similarly, if the site $n$ is occupied by a quencher, 
inverse transition is possible for the state, 
$| \vec{r}_n, \bullet \rangle | \vec{r}_m, \phi \rangle$, 
if at least one neighboring site is vacant.   
$H_{\rm rc}$ describes the reaction from an occupied site $\vec{r}_n$ 
with the rate $k \left(\vec{r}_n \right)$, 
\cite{Doi,Kotomin,Peliti}
\begin{eqnarray}
H_{\rm rc} = - \sum_{n=1}^M k \left(\vec{r}_n \right) | \vec{r}_n, \bullet \rangle \langle \vec{r}_n, \bullet | .
\end{eqnarray}

In order to calculate the survival probability, 
Eq. (\ref{survivaldef}), from the configuration probability, 
it is convenient to introduce the generating function, \cite{vanKampen}
\begin{eqnarray}
G(x,t) &=& \sum_{N=0}^M \sum_{\{ \vec{r}_i\}} x^N P(\vec{r}_1, \vec{r}_2, \cdots , \vec{r}_N; t). 
\end{eqnarray}
The survival probability is obtained from, \cite{vanKampen}
\begin{eqnarray}
P_N (t) = e^{-t/\tau_0} \frac{1}{2 \pi i} \int d\, x \frac{1}{x^{N+1}} G (x, t) , 
\label{inversetf} 
\end{eqnarray} 
where the generating function is rewritten as, \cite{Nakazato}
\begin{eqnarray}
G(x,t) &=& \frac{1}{_MC_N} 
\prod_{\ell=1}^M \left( \langle  \vec{r}_\ell, \phi | + 
\sqrt{x}  \langle \vec{r}_\ell, \bullet |\right)
e^{Ht} 
\prod_{k=1}^M \left( | \vec{r}_k, \phi \rangle + 
\sqrt{x} |\vec{r}_k, \bullet \rangle 
\right). 
\label{generate}
\end{eqnarray}
Eq. (\ref{inversetf}) with Eq. (\ref{generate}) generalizes  
Eq. (\ref{staticinv1}) by including the effect of diffusion with 
excluded volume interactions. 

In the thermodynamic limit in which $M$ tends to infinity with 
the fraction of quenchers being fixed, 
$c=N/M$, 
we can apply a saddle point method 
to Eq. (\ref{inversetf}) as we have done to obtain 
Eq. (\ref{staticinv3}). 
Originally, the method is  
introduced by Nakazato and Kitahara for the 
calculation of tracer diffusion constant of correlated random walk. \cite{Nakazato} 
The same result as theirs can be obtained by a different method. 
\cite{Benichou}
The results of Nakazato and Kitahara is also confirmed 
by numerical simulation in 2 and 3 dimensional systems.  
%revised
\cite{Suzuki,Okamoto07,vanBeijeren85}
The method is based on the fact 
that the number of diffusing quenchers is conserved. 
In our case, 
the number of quenchers is conserved for the quencher configurations which survive reaction,  
and the correlated random walks are performed by exactly $N$ quenchers.
By applying a saddle point method, 
Eq. (\ref{inversetf}) becomes, 
\begin{eqnarray}
P_N (t) =e^{-t/\tau_0} \left( \prod_{\ell=1}^M \langle \vec{r}_\ell, \phi | \right) 
\exp \left( \tilde{H} t \right)  \left( \prod_{\ell=1}^M | \vec{r}_\ell, \phi \rangle \right) , 
\label{pop_1}
\end{eqnarray}
where $\tilde{H}= \exp \left( - \theta^* S \right) H  \exp \left( \theta^* S \right)$, 
$S\equiv \sum_{\ell=1}^M \left(  |\vec{r}_\ell, \bullet \rangle \langle  \vec{r}_\ell, \phi |
- |\vec{r}_\ell, \phi \rangle  \langle  \vec{r}_\ell, \bullet | 
\right) 
$, and 
$x=\tan^2 \theta$, 
with $\tan \theta^* = \sqrt{c/(1-c)}$. 
$\tilde{H}$ is obtained as $\tilde{H}= \tilde{H}_0 + \tilde{H}_1$, where, 
\begin{eqnarray}
\tilde{H}_0 = H_{\rm w} - \sum_{n=1}^M k\left(\vec{r}_n\right) 
\left[ 
(1-c)  | \vec{r}_n, \bullet \rangle \langle \vec{r}_n, \bullet |+ 
c | \vec{r}_n, \phi \rangle \langle \vec{r}_n, \phi | 
\right]
\end{eqnarray}
and 
\begin{eqnarray}
\tilde{H}_1 =  - \sum_{n=1}^M k\left(\vec{r}_n\right) \sqrt{c(1-c)} 
\left( 
| \vec{r}_n, \bullet \rangle \langle \vec{r}_n, \phi |+ 
| \vec{r}_n, \phi \rangle \langle \vec{r}_n, \bullet | 
\right) .
\label{defH1}
\end{eqnarray}
By making time differentiation of Eq. (\ref{pop_1}), 
we obtain the time evolution equation for the survival probability, 
\begin{eqnarray}
\frac{\partial}{\partial t} P_N (t) = - \frac{1}{\tau_0} P_N(t) 
 -  c \kappa P_N (t) - 
\sqrt{c ( 1-c)} \sum_{j=1}^M k \left(\vec{r}_j \right) q \left( \vec{r}_j, t \right), 
\label{tevop}
\end{eqnarray}
where the sum of reaction rates is defined as, 
\begin{align}
\kappa = \sum_{j=1}^M k \left(\vec{r}_j \right).
\label{kappa} 
\end{align}
$q \left( \vec{r}_j, t \right)$ is 
given by, 
\begin{eqnarray}
q \left( \vec{r}_j, t \right) \equiv e^{-t/\tau_0}
\left( \prod_{\ell=1}^{M'} \langle \vec{r}_\ell, \phi | \langle \vec{r}_j, \bullet| \right) 
\exp \left( \tilde{H} t \right) 
\left( \prod_{\ell=1}^{M} | \vec{r}_\ell, \phi \rangle\right) ,
\label{defq}
\end{eqnarray}
where $M'$ denotes that the site $\vec{r}_j$ is excluded in the product. 
The initial condition of Eq. (\ref{tevop}) is $P(0)=1$. 
After Laplace transformation, eq(\ref{tevop}) leads to, 
\begin{eqnarray}
\hat{P}_N (s) = \frac{1}{s+1/\tau_0 + \sum_{j=1}^M c k \left(\vec{r}_j \right) }
\left[ 1 - \sqrt{c ( 1-c)} \sum_{j=1}^M k \left(\vec{r}_j \right) \hat{q} \left( \vec{r}_j, s \right) \right],  
\label{psq}
\end{eqnarray}
where 
$\hat{q} \left( \vec{r}_j, s \right) = \int_0^\infty \,dt \exp (-st) q \left( \vec{r}_j, t \right)$. 

$q \left( \vec{r}_j, t \right)$ is calculated by the perturbation expansion of 
$\exp \left( \tilde{H} t\right)$. 
$\tilde{H}_1$ is taken as the perturbation term. 
$\tilde{H}_0$ conserves the number of quenchers. 
On the other hand, 
$\tilde{H}_1$ defined by Eq. (\ref{defH1}) changes the number of quenchers by the amount of one 
and only the odd powers of $\tilde{H}_1$ contribute in the perturbation expansion. 
The expansion parameter is proportional to $c(1-c)$ instead of $\sqrt{c(1-c)}$ 
given in the definition of $\tilde{H}_1$.

A simple expression is obtained by the P\'{a}de approximation, 
\begin{eqnarray}
\hat{q} \left( \vec{r}_j, s \right) &=& - 
\frac{\displaystyle \sum_{\ell=1}^M \hat{G} \left(\vec{r}_j,\vec{r}_\ell, s \right) k\left( \vec{r}_\ell \right) \sqrt{c(1-c)} 
\frac{1}{s+1/\tau_0 +c \kappa}}
{\displaystyle 1-c(1-c) \sum_{v=1}^M \sum_{w=1}^M 
\frac{k\left( \vec{r}_v \right) \hat{G}\left(\vec{r}_v, \vec{r}_w,s \right)  k\left( \vec{r}_w \right)}
{s+1/\tau_0 +c \kappa}} ,
\label{solq}
\end{eqnarray}
where 
$\kappa$ is defined by Eq. (\ref{kappa}) and 
$\hat{G} \left(\vec{r}_i, \vec{r}_j, s \right)$ is the Laplace transform of, 
\begin{eqnarray}
G \left(\vec{r}_i, \vec{r}_j, t \right)\equiv e^{-t/\tau_0}
 \left( \prod_{\ell=1}^{M'} \langle \vec{r}_\ell, \phi | \langle \vec{r}_i, \bullet| \right) 
\exp \left( \tilde{H}_0 t \right) 
\left( \prod_{\ell=1}^{M'} | \vec{r}_\ell, \phi \rangle | \vec{r}_j, \bullet \rangle \right) .
\label{defG}
\end{eqnarray}

In the P\'{a}de approximation, 
higher order Green's functions such as, 
$\hat{G} \left(\vec{r}_1,\vec{r}_2, \vec{r}_3| \vec{r}_4, \vec{r}_5,\vec{r}_6, s \right)$ 
defined similarly to 
Eq. (\ref{defG}) are ignored. 
As stated before, odd powers of $\tilde{H}_1$ should be 
left in the perturbation expansion. 
Since $\tilde{H}_1$ is a reaction term as shown in Eq. (\ref{defH1}), 
the results gives minus contribution to the perturbation expansion. 
If we denote the complete solution including higher order Green's functions by $q_T \left( \vec{r}_j, s \right)$, 
the approximate solution $q \left( \vec{r}_j, s \right)$ obeys, 
$q \left( \vec{r}_j, s \right) \geq q_T \left( \vec{r}_j, s \right)>0$. 
By combining this inequality with Eq. (\ref{psq}), 
we find that the approximate expression is a lower bound of $P_N (t)$.

By making time differentiation of Eq. (\ref{defG}), we obtain, 
\begin{eqnarray}
\frac{\partial}{\partial t} G\left(\vec{r}_i, \vec{r}_j, t \right) 
&=& 
- \frac{1}{\tau_0} G\left(\vec{r}_i, \vec{r}_j, t \right) +
{\cal L} G\left(\vec{r}_i, \vec{r}_j, t \right) 
\nonumber \\
& & \mbox{ }-  (1-2c) k \left(\vec{r}_i \right) G\left(\vec{r}_i, \vec{r}_j, t \right) 
- c \kappa G\left(\vec{r}_i, \vec{r}_j, t \right), 
\label{closedG}
\end{eqnarray}
where the initial condition is given by $G\left(\vec{r}_i, \vec{r}_j, 0 \right) = \delta_{i,j}$. 
${\cal L}$ represents the operator describing hopping transitions, 
\begin{align}
{\cal L}  G\left(\vec{r}_i, \vec{r}_j, t \right) 
= \sum_{k=1}^{2d} \Gamma/(2d)  
\left[ G\left(\vec{b}_k+\vec{r}_i, \vec{r}_j, t \right) - G\left(\vec{r}_i, \vec{r}_j, t \right) \right] ,
\label{hopping}
\end{align} 
where $d$ is the dimensionality of hypercubic lattice. 
$\vec{b}_k+\vec{r}_i$ denotes a nearest neighbor of the site $\vec{r}_i$ 
and the sum is taken over all nearest neighbor sites.

By introducing Eq. (\ref{solq}) into Eq. (\ref{psq}) in the Laplace domain, 
the Laplace transform of the survival probability is expressed as, 
\begin{eqnarray}
\hat{P}_N (s) = \frac{1}{\displaystyle s+ 1/\tau_0 + c\kappa - 
c (1-c) \sum_{v=1}^M \sum_{w=1}^M 
k\left( \vec{r}_v \right) \hat{G}\left(\vec{r}_v, \vec{r}_w,s \right)  k\left( \vec{r}_w \right)} .
\label{LaplaceP}
\end{eqnarray}
We can obtain the survival probability, $P_N (t)$,  
by introducing the solution of Eq. (\ref{closedG}) into Eq. (\ref{LaplaceP}) 
and making the inverse Laplace transformation.  
The term with $c(1-c)$ represents the effect of correlated diffusion, 
which vanishes in the dilute limit, $c \rightarrow 0$. 
In the opposite limit of $c \rightarrow 1$, 
the factor $c(1-c)$ again vanishes corresponding to the absence of diffusion 
since every site is occupied by a quencher. 
In both limits, the survival probability is given by, 
\begin{eqnarray}
P_N (t) = \exp \left( -t/\tau_0 - c \kappa t \right) .  
\label{highlowconc}
\end{eqnarray}
Eq. (\ref{highlowconc}) reproduces Eq. (\ref{staticfamiliar}) 
obtained for the static quenching when all  sites 
are occupied by quenchers. 
In the dilute limit, $c\rightarrow 0$, 
Eq. (\ref{highlowconc}) is also consistent with 
the known result of static quenching,  
Eq. (\ref{staticknown}),  
when the reaction rate is small. 
In the intermediate concentration, the survival probability is influenced by 
$G\left(\vec{r}_i, \vec{r}_j, t \right)$ defined by 
the probability of finding a quencher at position $\vec{r}_i$ at time $t$ 
when it starts from $\vec{r}_j$ under the 
prohibition of double occupancy of a site.

%%%%%%%%%%%%%%%%%%%%%%%%%%%%%%%%%%%%%%%%%%%%%%%%%%%%%%%%%%%%%%%%%%%%%%
% Localized sink %%%%%%%%%%%%%%%%%%%%%%%%%%%%%%%%%%%%%%%%%%%%%%%%%%%%%%%%%%%%%%
%%%%%%%%%%%%%%%%%%%%%%%%%%%%%%%%%%%%%%%%%%%%%%%%%%%%%%%%%%%%%%%%%%%%%%
%\setcounter{equation}{0}
\section{Localized reactions}
\vspace{0.5cm}

When reaction takes place 
only at a target site $\vec{r}_R$, $k \left(\vec{r}_\ell \right) = k_0 \delta_{\vec{r}_\ell, \vec{r}_R}$, 
the Laplace transform of the survival probability is expressed as, 
\begin{eqnarray}
\hat{P}_N (s) = \frac{1}{\displaystyle s+ 1/\tau_0 + c  k_0 - 
c (1-c) 
k_0 \hat{G}\left(\vec{r}_R, \vec{r}_R,s \right)  k_0} .
\label{LaplacePdl}
\end{eqnarray}
In the absence of diffusion and natural decay, 
Eqs. (\ref{closedG}) and (\ref{LaplacePdl}) 
reproduce Eq. (\ref{staticLaplace0}) derived by assuming 
the static quenching from the beginning.

In the presence of diffusion, 
Eq. (\ref{LaplacePdl}) represents the approximate solution 
which interpolates between solutions in low and high limits of quencher concentrations. 
By substituting 
$k \left(\vec{r}_\ell \right) = k_0 \delta_{\vec{r}_\ell, \vec{r}_R}$, 
the solution of Eq. (\ref{closedG}) in the Laplace space can be expressed as
\begin{align}
\hat{G} (\vec{r}_R,\vec{r}_R,s) = \frac{\hat{G}_0(\vec{r}_R,\vec{r}_R,z)}
{1+ \left(1-2c \right) k_0 \hat{G}_0(\vec{r}_R,\vec{r}_R,z)} . 
\label{sol2}
\end{align}
$\hat{G}_0(\vec{r}_i,\vec{r}_j,s)$ is the Laplace transform of the 
Green's function satisfying, 
\begin{align}
\frac{\partial G_0 \left(\vec{r}_i, \vec{r}_j, t \right)}{\partial t} 
= {\cal L} G_0 \left(\vec{r}_i, \vec{r}_j, t \right) + \delta_{\vec{r}_i, \vec{r}_j} \delta (t),   
\label{sv3} 
\end{align}
where ${\cal L}$ represents the operator describing hopping transitions   
given by Eq. (\ref{hopping}). 
$\hat{G}_0(\vec{r}_R,\vec{r}_R,z)$ in Eq. (\ref{sol2}) is given in terms of 
the Green's function for free random walks, 
$\hat{G}_0(\vec{r}_R,\vec{r}_R,s)$, 
but the Laplace variable is modified 
as a result of the excluded volume interactions among quenchers and 
expressed in terms of 
the initial concentration of quenchers and the 
reaction rate, $c k_0$,  
\begin{align}
z=s+(1/\tau_0)+c k_0. 
\label{eq:z}
\end{align}

When quenchers can migrate on all lattice sites including the target site, 
the Laplace transform of the 
Green's function in the absence of site blocking effects and the reaction can be written as   
\begin{align}
\hat{G}_0 \left( \vec{r}_R,\vec{r}_R, s \right) 
&= 
\frac{1- \hat{\psi} (s)}{s} U(s) 
\label{sv4_1}
\end{align}
in terms of the Lattice Green's function $U(s)$ defined by
\begin{align}
U(s) &= \frac{1}{(2\pi)^d} \int \cdots \int_{-\pi}^{\pi} d^d \vec{k}
\frac{1}{1 - \hat{\psi} (s) \tilde{\lambda} (\vec{k})}, 
\label{latticeU}
\end{align}
where $\hat{\psi} (s)=\Gamma/(s+ \Gamma)$, 
and the structure factor is defined by 
$\tilde{\lambda} (\vec{k}) \equiv \frac{1}{2d} \sum_{j=1}^{2d} \cos \left(\vec{k} \cdot \vec{b}_j/b \right)$.  
$b$ denotes the lattice spacing. 
Eq. (\ref{LaplacePdl}) can be rewritten as, 
\begin{align}
\hat{P}_N (s) &= 
\frac{1}
{\displaystyle 
s+ (1/\tau_0) + 
c\frac{1}{1/k_0+(1-c)/\left( 1/\hat{G}_0-c k_0\right)}},
\label{sol1}
\end{align}
where we use the abbreviation, $\hat{G}_0=\hat{G}_0(\vec{r}_R,\vec{r}_R,z)$ 
and $z$ is defined by Eq.(\ref{eq:z}). 
Eq. (\ref{sol1}) is one of the most important results of this paper.  

Eq. (\ref{sol1}) is simplified in the Smoluchowski limit 
which is given by $k_0 \rightarrow \infty$. 
In order to obtain the limit, we rewrite Eq. (\ref{sol1}) as
\begin{align}
\hat{P}_N (s) &= 
\frac{1}
{\displaystyle 
s+ (1/\tau_0) + 
c\frac{1}{1/k_0+(1-c)U(z)/\left(s+(1/\tau_0)+\Gamma+c k_0 (1-U(z))\right)}}. 
\label{sol_Sm}
\end{align}
By introducing the explicit expression of $U(z)$, 
we find, 
\begin{align}
\lim_{c k_0 \rightarrow \infty} U(z)=1 \mbox{ and } \lim_{c k_0 \rightarrow \infty} c k_0 (1-U(z)) =0 ,
\label{sol_Sm1}
\end{align}
for any spatial dimension. 
In the limit of $k_0 \rightarrow \infty$ 
(hopping-controlled limit), 
Eq. (\ref{sol_Sm}) is simplified into 
\begin{align}
 \hat{P}_N (s) &=\frac{1-c}{s+1/\tau_0+c \Gamma}.  
\label{solhoppingl}
\end{align}
Subsequent inverse Laplace transformation yields a single exponential decay, 
\begin{align}
P_N (t) = (1-c) \exp \left[ - \left(1/\tau_0 + c \Gamma \right) t \right] . 
\label{soldifflattice}
\end{align}
In the limit of $c=1$, 
the reaction site is occupied by a quencher at the initial time and 
the reaction takes place immediately in the limit of $k_0 \rightarrow \infty$. 
The probability that the reaction site is not occupied by 
a quencher is given by 
$1-c$ and the reaction takes place with the rate $c \Gamma$ which is proportional 
to both the hopping rate and the quencher concentration. 
In the hopping-controlled limit, 
Eq. (\ref{soldifflattice}) is a lower bound of the survival probability.

In the Smoluchowski limit of $k_0 \rightarrow \infty$ for a localized sink in 1 dimensional systems, 
the survival probability shows non-exponential decay if double occupancy of sites 
is allowed. 
The non-exponential decay in the presence of 
site blocking effects is also predicted by some theories. 
\cite{SzaboPRL,Bramson,Arora,Burlatsky,Bhatia,Sokolov} 
However, our approximate results predict 
the exponential decay in the limit of $k_0 \rightarrow \infty$. 
Since our derivation involves the 
steepest descent approximation of Nakazato-Kitahara's theory and 
the P\'{a}de approximation of perturbation expansion, 
there should be a certain limitations on our theory. 
In the absence of diffusion 
our theory predicts 
the exact results of static quenching 
for localized reactions,   
regardless of the dimensionality of the systems. 
However, in the presence of diffusion,  
it gives only 
a lower bound of the survival probability. 
The accuracy of the approximation is worse 
in the limit of $k_0 \rightarrow \infty$ in the presence of diffusion. 
The accuracy also depends on the dimensionality of the systems. 
We conjecture that the perturbation term which appeared by applying Nakazato-Kitahara's theory 
is large in the limit of $k_0 \rightarrow \infty$ 
in the presence of diffusion in one dimensional systems. 

%%% Adjoint method
%%%%%%%%%%%%%%%%%%%%%%%%%%%%%%%%%%%%%%%%%%%%%%%%%%%%%%%%%%%%%%%%%%%%%%
% Simplification by adjoint equation %%%%%%%%%%%%%%%%%%%%%%%%%%%%%%%%%%%%%%%%%%%%%%%%%%%%%%%%%%%%%%
%%%%%%%%%%%%%%%%%%%%%%%%%%%%%%%%%%%%%%%%%%%%%%%%%%%%%%%%%%%%%%%%%%%%%%
%\setcounter{equation}{0}
\section{Simplification by adjoint equation}
\vspace{0.5cm}

When
the Green's function is not known, 
it is convenient to define 
the pair survival probability, 
\begin{align}
f\left( \vec{r}_\ell, t\right) =  \sum_{i=1}^M   G\left(\vec{r}_i, \vec{r}_\ell, t \right) , 
\end{align}
which describes the survival probability of a pair whose initial 
separation is given by a vector $\vec{r}_\ell$. 
From the equation for $f\left( \vec{r}_\ell, t\right)$, 
the bulk survival probability 
can be obtained without knowing the Green's function. 
As shown below, 
the equation for $f\left( \vec{r}_\ell, t\right)$ is simpler than 
that for the Green's function. 
The initial condition is given by, 
\begin{align}
f\left( \vec{r}_\ell, t=0\right) = 1 . 
\end{align}
$f \left( r_\ell, t \right)$ satisfies the time evolution equation with the diffusional operator 
${\cal L}^\dagger$ adjoint with ${\cal L}$, 
\begin{align}
\frac{\partial}{\partial t} f\left(\vec{r}_\ell, t \right) &= 
- \frac{1}{\tau_0} f\left(\vec{r}_\ell,  t \right) + {\cal L}^\dagger f \left(\vec{r}_\ell, t \right) 
-\left(1-2c \right)  k \left(\vec{r}_\ell \right) f\left(\vec{r}_\ell,  t \right) 
-  c  \kappa f\left(\vec{r}_\ell,  t \right) .  
\label{closedf}
\end{align}
This is a generalization of the time evolution equation of the pair survival probability derived by 
Sano and Tachiya. \cite{Sano}
In the absence of potential, 
${\cal L}^\dagger$ and ${\cal L}$ are equal, 
$
{\cal L}^\dagger = {\cal L}
$.  
From Eq. (\ref{closedG}), we obtain the following relation,  
\begin{align}
\frac{\partial}{\partial t} 
\sum_{i=1}^M  \sum_{j=1}^M  
G\left(\vec{r}_i, \vec{r}_j,t \right)  k\left( \vec{r}_j \right)
&= -(1 -2c) 
\sum_{i=1}^M  \sum_{j=1}^M 
k\left( \vec{r}_i \right) G\left(\vec{r}_i, \vec{r}_j,t \right)  k\left( \vec{r}_j \right) 
\nonumber \\
&-c \kappa 
\sum_{i=1}^M  \sum_{j=1}^M
G\left(\vec{r}_i, \vec{r}_j,t \right)  k\left( \vec{r}_j \right),
\end{align}
and after the Laplace transformation it leads to 
\begin{align}
\sum_{i=1}^M  \sum_{j=1}^M  
k\left( \vec{r}_i \right) \hat{G}\left(\vec{r}_i, \vec{r}_j,s \right)  k\left( \vec{r}_j \right)
= 
\frac{\kappa- (s+c \kappa) 
\sum_{i=1}^M  \sum_{j=1}^M 
\hat{G}\left(\vec{r}_i, \vec{r}_j,s \right)  k\left( \vec{r}_j \right)
}{1-2c}  . 
\label{relationf}
\end{align}
By substituting Eq. (\ref{relationf}), Eq. (\ref{LaplaceP}) can be rewritten as, 
\begin{align}
\hat{P}_N (s) = \frac{1}{
\displaystyle s+1/\tau_0+\left( 
-c^2 \kappa +c(1 -c) z
\sum_{\ell=1}^M   k (\vec{r}_\ell) \hat{f} \left( \vec{r}_\ell, s \right)
\right)/(1-2c)
}, 
\label{adjoint1}
\end{align}
where $z$ is given by Eq. (\ref{eq:z}).
This is a generalization of the equation for the survival probability derived by Tachiya,  
by taking into account the site blocking effects. \cite{Tachiya83} 
The expression for the reaction rate is known 
for localized reactions, 
which leads to, \cite{Rice,Tachiya83,Szabo89} 
\begin{align}
z
\sum_{\ell=1}^M k (\vec{r}_\ell) \hat{f} \left( \vec{r}_\ell, s \right) =
\frac{1}{
\displaystyle 
(1/k_0)+ 
(1-2c)\hat{G}_0(\vec{r}_R,\vec{r}_R,z) 
} .
\label{adjointlcl}
\end{align}
By substituting Eq. (\ref{adjointlcl}) into Eq. (\ref{adjoint1}), 
we reproduce 
Eq. (\ref{LaplacePdl}) with Eq. (\ref{sol2}). 
For localized reactions, 
$G\left(\vec{r}_1, \vec{r}_2,t \right) $ is known and 
the adjoint equation may not be needed. 
However, for long-range reactions, 
calculation of
$\hat{f} \left( \vec{r}_\ell, s \right)$ using  the equilibrium initial condition 
can be easier than that of $\hat{G}(\vec{r}_R,\vec{r}_R,s)$ 
using the initial condition expressed by  
Kronecker's delta.

%%%%%%%%%%%%%%%%%%%%%%%%%%%%%%%%%%%%%%%%%%%%%%%%%%%%%%%%%%%%%%%%%%%%%%
% Stern-Volmer law; Steady-state fluorescence %%%%%%%%%%%%%%%%%%%%%%%%%%%%%%%%%%%%%%%%%%%%%%%%%%%%%%%%%%%%%%
%%%%%%%%%%%%%%%%%%%%%%%%%%%%%%%%%%%%%%%%%%%%%%%%%%%%%%%%%%%%%%%%%%%%%%
%\setcounter{equation}{0}
\section{Stern-Volmer law}
\vspace{0.5cm}

In this section, we study the site blocking effects of diffusion on Stern-Volmer law. 
A Stern-Volmer plot is obtained from the fluorescence intensity at different quencher concentrations. 
The relative fluorescence intensity against $\eta_0$ defined 
in the absence of quencher is given by, \cite{Inokuti,Rice}
\begin{eqnarray}
\frac{\eta}{\eta_0} 
= \frac{\int_0^\infty dt P_N (t)}{\int_0^\infty dt P_0 (t)}
= \frac{1}{\tau_0} \hat{P}_N (s=0). 
\end{eqnarray} 

By substituting Eq. (\ref{LaplaceP}),
$\eta /\eta_0$ 
is obtained as,  
\begin{eqnarray}
\frac{\eta}{\eta_0}  =\frac{1}{\tau_0}  
\frac{1}{\displaystyle 1/\tau_0 + c \kappa - 
c (1-c) \sum_{v=1}^M \sum_{w=1}^M 
k\left( \vec{r}_v \right) \hat{G}\left(\vec{r}_v, \vec{r}_w,0 \right)  k\left( \vec{r}_w \right)} .
\label{SVP}
\end{eqnarray}
In the Stern-Volmer plot, $\eta_0/\eta-1$ is plotted against the concentration, $c$, 
\begin{eqnarray}
\eta_0/\eta -1 = c \kappa \tau_0 - 
c (1-c) \tau_0 \sum_{v=1}^M \sum_{w=1}^M 
k\left( \vec{r}_v \right) \hat{G}\left(\vec{r}_v, \vec{r}_w,0 \right)  k\left( \vec{r}_w \right) ,
\label{SVt}
\end{eqnarray}
It increases linearly with $c$ when quenchers are dilute. 
Deviation from linear concentration dependence of $\eta_0/\eta-1$ is 
theoretically obtained by solving the equation for $\hat{G}\left(\vec{r}_1, \vec{r}_2,s \right)$ 
given by the Laplace transform of Eq. (\ref{closedG}).

For the localized reactions, 
$k \left(\vec{r}_\ell \right) = k_0 \delta_{\vec{r}_\ell, \vec{r}_R}$, 
we obtain the following equation 
by substituting $s \rightarrow 0$ limit of Eq. (\ref{sol2}) into Eq. (\ref{SVt}), 
\begin{align}
\eta_0/\eta -1 &= \frac{c \tau_0}
{\displaystyle 1/k_0+
\frac{1-c}
{1/\hat{G}_0(\vec{r}_R,\vec{r}_R,z_0) -ck_0}}  
\label{STlocalg}
\\
&= \frac{c k_0 \tau_0}{
\displaystyle 
1+ 
\frac{\left( 1-c \right) U(z_0)}
{c \left[1 - U(z_0) \right] + 
(1/\tau_0+\Gamma)/k_0 
}},  
\label{SVtcslocal1_1}
\end{align}
where $s \rightarrow 0$ limit of $z$ is introduced, 
\begin{align}
z_0=1/\tau_0+ ck_0 . 
\label{eq:z0}
\end{align}
The Laplace variable given by Eq. (\ref{eq:z0}) 
includes the effect of initial concentration of quenchers and the reaction rate. 
This is a signature of  
the excluded volume interactions among quenchers. 
Eq. (\ref{STlocalg}) is positive since we can prove 
\begin{align}
1 > ck_0\hat{G}_0(\vec{r}_R,\vec{r}_R,z_0) >0 ,
\label{conditionl1}
\end{align}
as shown in appendix. 
Eq. (\ref{STlocalg}) is an important result of this paper.

In the static limit, 
Eq. (\ref{STlocalg}) reduces to the following equation  
by substituting  
$\hat{G}_0(\vec{r}_R,\vec{r}_R,z_0)=1/z_0$, 
\begin{align}
\eta_0/\eta -1  = \frac{c}{1-c+1/(\tau_0 k_0)}
. 
\label{SVstatic}
\end{align}
This is the exact result. 

In the limit of $k_0 \rightarrow \infty$, 
the result in the hopping-controlled limit is obtained from Eq. (\ref{SVtcslocal1_1}) as,   
\begin{align}
\eta_0/\eta -1  &=
c \frac{1 + \Gamma \tau_0}{1-c} . 
\label{svdiff}
\end{align}
The above expression shows 
that $\eta_0/\eta -1$ increases linearly with increasing the hopping frequency, $\Gamma$,  
for any concentration. 
In the limit of $c=1$, 
the target site is occupied by a quencher at the initial time and reaction takes place 
with probability $1$ when $k_0 \rightarrow \infty$. 
In the opposite limit of $c\rightarrow 0$, 
$\eta_0/\eta-1$ is proportional to the concentration $c$. 
By time integration, we can show that Eq. (\ref{svdiff}) is consistent with Eq. (\ref{soldifflattice}).

For various lattices, the lattice Green's function, Eq. (\ref{latticeU}), is known. 
As an example,  we consider 
random walks on the BCC lattice. 
The reaction takes place at the site $\vec{r}_R$ with the rate $k_0$. 
Without loss of generality, 
the target site $\vec{r}_R$ can be taken at the origin of the lattice. 
Quenchers perform random walks on the lattice including the origin, and 
each site can be occupied at most by a single quencher. 
The lattice Green's function is known, \cite{Hughes}
\begin{eqnarray}
\hat{G}_0 \left( \vec{r}_R,\vec{r}_R, s \right) = 
\left\{_2F_1 \left(\frac{1}{4}, \frac{1}{4}; 1; \left(\Gamma/(s+\Gamma) \right)^2 \right) \right\}^2 
/(s+\Gamma).
\label{sv5}
\end{eqnarray}
When $\Gamma < z_0$, 
we can approximate 
$_2F_1 \left(\frac{1}{4}, \frac{1}{4}; 1; \xi^2 \right) \sim 1 + \xi^2/16$, for $\xi \rightarrow 0$ 
\cite{Abramowitz} in  
Eq. (\ref{sv5}), 
and 
Eq. (\ref{STlocalg}) is expressed as, 
\begin{align}
\eta_0/\eta -1&=
\frac{c\tau_0}{
\displaystyle 
\frac{1}{k_0} + 
\frac{1-c}{
1/\tau_0 + \Gamma -
(ck_0/8) 
\Gamma^2/\left(z_0 + \Gamma 
\right)^2
}
} . 
\label{svpdiff}
\end{align}
In the case of $c k_0 > \Gamma/8$, Eq. (\ref{svpdiff}) is further simplified as, 
\begin{align}
\eta_0/\eta -1&=
\frac{c \tau_0}{
\displaystyle 
\frac{1}{k_0} + 
\frac{1-c}{
1/\tau_0 + \Gamma}
} . 
\label{svpdiffs}
\end{align}
Eq. (\ref{svpdiffs}) is the result valid irrespective of the lattice structure  
since it can be derived by introducing the approximation,
$\hat{G}_0 \left( \vec{r}_R,\vec{r}_R, z_0 \right) \sim 1/(z_0+\Gamma)$ 
which is  
valid when $\Gamma < z_0$, into 
Eq. (\ref{STlocalg}). 
In the reaction-controlled or static limit, we obtain Eq. (\ref{SVstatic}), whereas, 
in the hopping-controlled limit, Eq. (\ref{svdiff}) is derived. 
Eq. (\ref{svpdiffs}) interpolates between the  static and hopping-controlled limits.

%========================================
% FIGURE
%========================================
\begin{figure}[htbp]%[p]%
\centerline{\includegraphics[width=0.6\columnwidth]{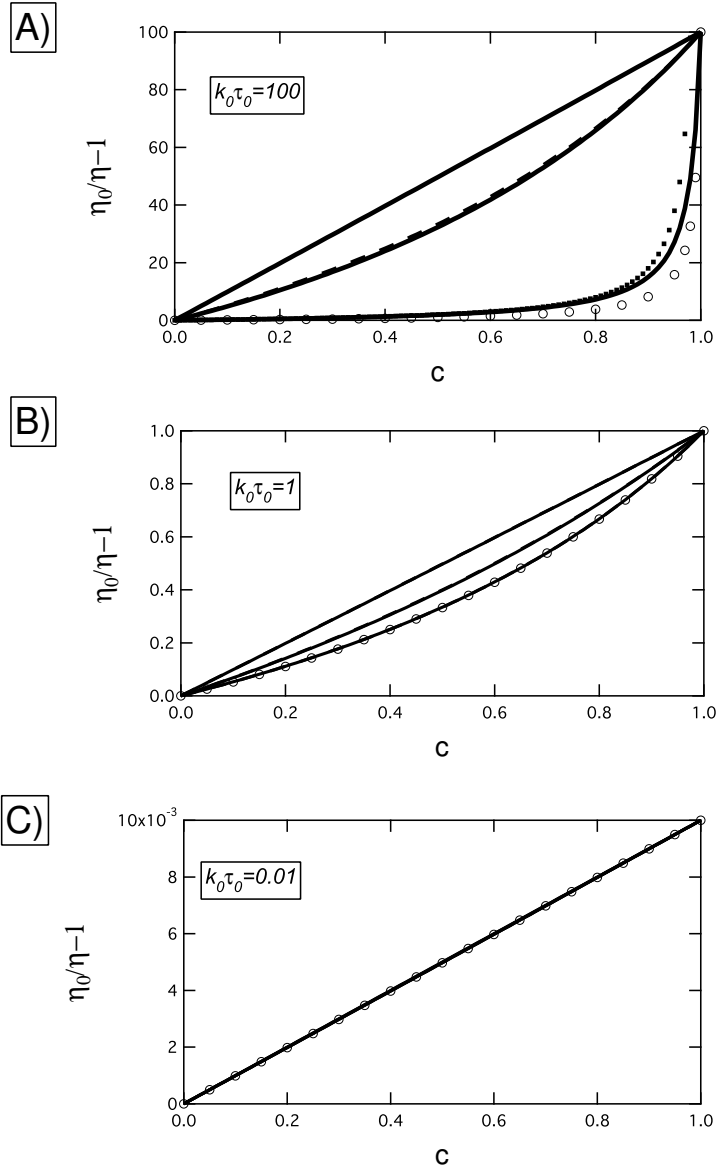}}
\caption{
$\eta_0/\eta-1$ against concentration $c$. 
A) $k_0 \tau_0 =100$; B) $ k_0 \tau_0=1$; C) $ k_0 \tau_0=0.01$. 
In all panels, curves correspond to $ \Gamma/k_0=100$, $ \Gamma/k_0=1$, and $ \Gamma/k_0=0.01$ 
from top to bottom. 
The thick solid lines indicate the general results 
 in the presence of site blocking effects, Eq. (\ref{STlocalg}) with Eq. (\ref{sv5}). 
 In C) they overlap. 
A dashed line in A) indicates the approximate result of Eq. (\ref{svpdiffs}).  
The other dashed lines are not visible, 
since they overlap with the solid lines. 
A dotted line in A) represents the results of hopping-controlled limit, Eq. (\ref{svdiff}), 
which is valid for $k_0 \rightarrow \infty$. 
Circles indicate the static solution, Eq. (\ref{SVstatic}). 
}
\label{fig:svBCC1}
\end{figure} 
%========================================
% FIGURE
%========================================
In Fig. \ref{fig:svBCC1}, the general results of Eq. (\ref{STlocalg}) with Eq. (\ref{sv5}) are plotted for various values of 
$k_0$ and $\Gamma$. 
The simplified solutions of Eq. (\ref{svpdiff}) 
overlap with those of Eq. (\ref{STlocalg}) with Eq. (\ref{sv5}) in Fig. 1. 
The further simplified solutions of Eq. (\ref{svpdiffs}) are also shown. 
The results of Eq. (\ref{svpdiffs}) 
reproduce the general results except for the case of 
$k_0 \tau_0 =100$ and $\Gamma /k_0=1$ 
where a small deviation is found. 
The results indicate that 
although Eq. (\ref{svpdiffs}) is derived under the condition of $8 c k_0 > \Gamma$, 
it is applicable in practice over a wide range. 
The results  
in the hopping-controlled limit of $k_0 \rightarrow \infty$, Eq. (\ref{svdiff}), 
are also shown for comparison. 
The results in the static limit, Eq. (\ref{SVstatic}), 
give the lower bound of $\eta_0/\eta$ for a given value of $k_0 \tau_0$.

%%%%%%%%%%%%%%%%%%%%%%%%%%%%%%%%%%%%%%%%%%%%%%%%%%%%%%%%%%%%%%%%%%%%%%
% Decay kinetics %%%%%%%%%%%%%%%%%%%%%%%%%%%%%%%%%%%%%%%%%%%%%%%%%%%%%%%%%%%%%%
%%%%%%%%%%%%%%%%%%%%%%%%%%%%%%%%%%%%%%%%%%%%%%%%%%%%%%%%%%%%%%%%%%%%%%
%\setcounter{equation}{0}
\section{Decay kinetics}
\vspace{0.5cm}

For localized reactions, 
the Laplace transform of the survival probability is obtained from Eq. (\ref{sol1}).  
In BCC lattice 
the lattice Green function is given by Eq. (\ref{sv5}). 
Therefore, 
when $\Gamma < 1/\tau_0 +c k_0$, 
Eq. (\ref{sol1}) is expressed as, 
\begin{align}
\hat{P}_N (s) = \frac{1}
{\displaystyle 
s+\frac{1}{\tau_0}+
\frac{c}
{\displaystyle 
1/k_0+ (1-c)/
\left[s+1/\tau_0+\Gamma -\left(c k_0/8 \right) \Gamma^2/
\left(z + \Gamma
\right)^2
\right]
}
} , 
\label{soldifflatticep}
\end{align}
where $z$ is given by Eq. (\ref{eq:z}) 
and approximation of Eq. (\ref{sv5}) using 
$_2F_1 \left(\frac{1}{4}, \frac{1}{4}; 1; \xi^2 \right) \sim 1 + \xi^2/16$ as 
$\xi \rightarrow 0$ is introduced.  
In the case of $c k_0 > \Gamma/8$, Eq. (\ref{soldifflatticep}) is simplified as, 
\begin{align}
\hat{P}_N (s) = \frac{1}
{\displaystyle 
s+\frac{1}{\tau_0}+
\frac{c}
{\displaystyle 
1/k_0+ (1-c)/
\left(s+1/\tau_0+\Gamma \right)
}
} . 
\label{soldifflatticepl}
\end{align}
The inverse Laplace transformation of Eq. (\ref{soldifflatticepl}) is obtained as, 
\begin{align}
P_N (t) = \frac{
\exp (-t/\tau_0)}{s_+ - s_-}
\left[ 
(s_+ - c k_0) \exp\left( - s_- t \right) - (s_- - c k_0) \exp\left( - s_+ t \right)
\right]
,  
\label{solhoplatticet}
\end{align}
where 
\begin{align}
s_{\pm} = \frac{\Gamma + k_0 \pm 
\sqrt{\left(\Gamma+k_0\right)^2 - 4 c \Gamma k_0}}{2} . 
\label{hoproots}
\end{align}
Eq. (\ref{solhoplatticet}) together with Eq. (\ref{hoproots}) is the result independent of the lattice structures.
Eqs. (\ref{solhoplatticet}) and (\ref{hoproots}) are derived under the condition  
$c k_0 > \Gamma/8$. 
Accordingly,  
the accuracy of the approximation decreases by decreasing the quencher concentration. 
The result in the hopping-controlled limit of $k_0 \rightarrow \infty$ 
reproduces Eq. (\ref{soldifflattice}).

In the reaction-controlled limit, 
Eq. (\ref{soldifflatticep}) reduces to 
\begin{align}
\hat{P}_N (s) = \frac{1}
{\displaystyle 
s+\frac{1}{\tau_0}+
\frac{c}
{\displaystyle 
1/k_0+ (1-c)/
\left(s+1/\tau_0 \right)
}
} , 
\label{soldifflatticepl1}
\end{align}
and its inverse Laplace transform is given by Eq. (\ref{static0}) when $1/\tau_0=0$.

For comparison, 
we present the conventional solution of the survival probability  
for target problem when site blocking effects among quenchers is completely ignored, 
\cite{Tachiya83,Szabo89}
\begin{align}
P_N (t) = \exp \left( - c \int_0^t d t_1 k_{\rm cv} (t_1)
\right) , 
\label{cklattice1}
\end{align}
where the Laplace transform of $k (t)$ is obtained from, 
\begin{align}
s \hat{k}_{\rm cv} (s) =  \frac{k_0}{1 + k_0 \hat{G}_0 \left( \vec{r}_R,\vec{r}_R, s \right) } . 
\label{cklattice2}
\end{align}

%========================================
% FIGURE conventional
%========================================
\begin{figure}[htbp]%[p]
\centerline{\includegraphics[width=0.55\columnwidth]{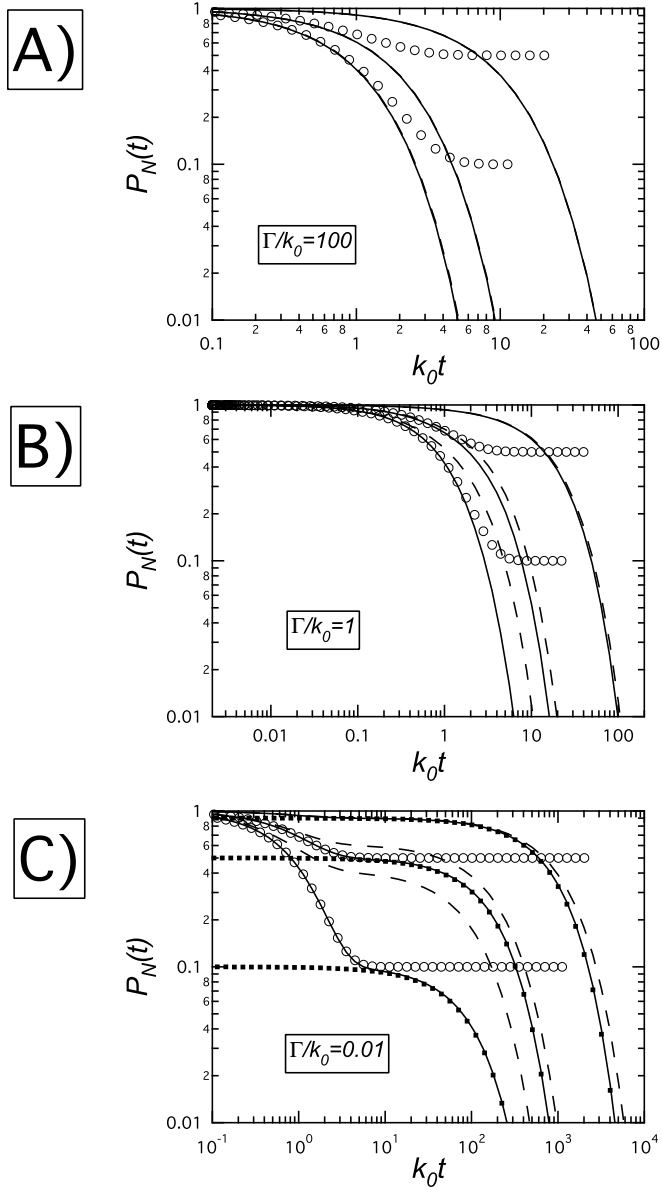}}
\caption{
The survival probability against normalized time, $k_0 t$.  $1/\tau_0=0$.  
A) $\Gamma/k_0 =100$; B) $\Gamma/k_0=1$; C) $\Gamma/k_0=0.01$. 
In all panels, curves correspond to $c=0.1$, $c=0.5$ and $c=0.9$ from right to left. 
The solid lines are obtained by the inverse Laplace transform of the 
general expression in the presence of site blocking effects, Eq. (\ref{sol1}) with Eq. (\ref{sv5}).
Dashed lines are obtained by the conventional expression, Eqs. (\ref{cklattice1}) and (\ref{cklattice2}). 
In A) the dashed lines are invisible because they overlap with the solid lines. 
Dotted lines in C) represent the solution in the hopping-controlled limit given by Eq. (\ref{soldifflattice}). 
Circles indicate the solution in the static limit given by Eq. (\ref{static0}).
}
\label{fig:BCCcnvk1}
\end{figure} 
%========================================
% FIGURE conventional
%========================================
In Fig. \ref{fig:BCCcnvk1}, 
the numerically obtained  
inverse Laplace transform of   
Eq. (\ref{sol1}) with Eq. (\ref{sv5}) is compared with 
the conventional solution, Eqs. (\ref{cklattice1}) and (\ref{cklattice2}). 
Excluded volume interaction is considered in Eq. (\ref{sol1}) with Eq. (\ref{sv5}), 
whereas it is ignored and each quencher is assumed to migrate independently 
in the conventional solution.  
In all cases, 
our results indicate that 
the survival probability in the presence of site blocking effects decays faster than that of the conventional solution where  
the excluded volume interaction is absent. 

In the dilute limit, $c \ll 1$, 
the difference between them is small regardless of the values of $\Gamma/k_0$. 
When the decay is mainly controlled by reaction, namely, $\Gamma/k_0 \gg 1$, 
the result of the conventional solution is close to that in the presence of site blocking effects 
even at high concentration of $c=0.9$. 
In this case, 
the excluded volume interaction is not important. 
As $\Gamma/k_0$ decreases, 
the deviation of the conventional solution from the results in the presence of site blocking effects increases at high concentrations. 

The initial decay of the solution of  Eq. (\ref{sol1}) with Eq. (\ref{sv5}) 
follows the results of static quenching from the uniform distribution, 
Eq. (\ref{static0}), over a longer period than that of the conventional solution given by 
Eqs. (\ref{cklattice1}) and (\ref{cklattice2}) 
in the cases of B) and C). 
In the initial time range, 
the decay of the survival probability takes place from the configuration where a quencher is 
initially located in the vicinity of the target. 
In the presence of site blocking effects,  
the migration of quenchers is suppressed and the initial decay follows the results of static quenching 
over a longer period than 
that derived under the assumption of free migration in the absence of site blocking. 
It should also be noticed that 
the survival probability obtained by assuming static quenching is the upper bound 
of that in the presence of diffusion, whereas the solution of Eq. (\ref{sol1}) with Eq. (\ref{sv5}) is the lower bound. 
The exact solution should lie between them.

When $\Gamma/k_0 \ll 1$, 
the initial time regime is approximated by the static quenching,  
and is followed by  
the hopping-controlled regime approximately described by Eq. (\ref{soldifflattice}) 
as shown in Fig. \ref{fig:BCCcnvk1}.

%%%%%%%%%%%%%%%%%%%%%%%%%%%%%%%%%%%%%%%%%%%%%%%%%%%%%%%%%%%%%%%%%%%%%%
% Conclusions %%%%%%%%%%%%%%%%%%%%%%%%%%%%%%%%%%%%%%%%%%%%%%%%%%%%%%%%%%%%%%
%%%%%%%%%%%%%%%%%%%%%%%%%%%%%%%%%%%%%%%%%%%%%%%%%%%%%%%%%%%%%%%%%%%%%%
%\setcounter{equation}{0}
\section{Conclusions}
\vspace{0.5cm}

We have investigated the target reaction problem 
in the presence of site blocking effects among quenchers. 
Quenchers migrate on any lattice sites until reaction takes place. 
Reaction rate depends on the distance between the quencher and 
the target. 
In the case of localized reactions, 
reaction takes place when a quencher comes to the target site. 
Once reaction occurs, the system becomes inert. 

The probability of reaction event is high if the excited target 
is initially surrounded by quenchers 
in close vicinity. 
As time proceeds, 
quencher configurations in which the quencher concentration near the excited target is low 
is more likely to survive  
than other configurations in the ensemble. 

In the conventional theory, 
excluded volume interactions among quenchers are ignored, {\it i.e.}, 
quenchers are regarded as independent of one another. 
However, 
quencher migration is hindered by the presence of other quenchers; 
the migration of a quencher is influenced by the time-dependent positions of other quenchers 
due to the site blocking effects.
We take into account the excluded volume interactions among quenchers 
by applying Nakazato-Kitahara's theory of vacancy-assisted diffusion. 
Our analytical solutions of the survival probability interpolate between those in two limits of 
low and high quencher concentrations and the approximation is good when the intrinsic reaction rate 
is low. 
When the intrinsic reaction rate is high and the condition for truncation of perturbation expansion, 
$c(1-c)k_0^2/(c k_0+ \Gamma)^2 <1$, is not satisfied, 
the higher order terms in the expansion is only partly taken into account by the P\'{a}de approximation. 
In other words, 
the higher order correlations 
originating from the diffusional collisions 
are not fully accounted for in the P\'{a}de approximation. 
In the presence of diffusion, 
our result is a lower bound of the survival probability. 
In the static limit, 
the exact results are reproduced from the P\'{a}de approximation.

The decay of the survival probability has been investigated for the target problem, 
where the target is excited by a pulse initially. 
The initial decay is well approximated by the static quenching. 
In particular, 
when the hopping frequency satisfies the relation, $\Gamma/k_0  < 1$, 
the initial decay of the survival probability at high quencher concentrations 
follows that of static quenching over a long period. 
The long time behavior of the general solution   
is approximated by the result in the hopping-controlled limit, Eq. (\ref{soldifflattice}), 
when $\Gamma/k_0 \ll 1$.

When the intrinsic reaction rate satisfies the relation, $\Gamma/k_0 \gg 1$,  
the conventional results in the absence of the site blocking effects 
reproduces those in the presence of the site blocking effects even at high quencher concentration of $c=0.9$.  
In the reaction-controlled limit, 
the excluded volume interaction among quenchers is not so important as that in the hopping-controlled limit.

According to the Brownian dynamic simulation, 
the survival probability in the presence of excluded volume interaction decays faster than that in its absence. 
\cite{Lee2000,Shin2003}
Similar enhancement of deactivation was also found by numerical simulation of random walk model on 
1 and 2 dimensional lattices. \cite{Arora,Burlatsky,Bhatia,Zumofen_EV}
Our results also suggest that the decay of the survival probability is accelerated by 
site blocking effects in the case of lattice random walk. 
The result can be understood as follows. 
Assume that there are $N$ quenchers on the lattice. As time proceeds, 
quenchers hop from site to site. Therefore, individual sites occupied by quenchers 
change with time. As long as the target site remains outside the sites occupied by quenchers, 
reaction does not occur. 
Once the sites occupied by quenchers include the target site, 
reaction occurs. 
In the absence of site blocking effects, 
different quenchers are allowed to occupy the same site. 
Therefore, in this case the number of the sites occupied by quenchers is generally less than $N$. 
In the presence of site blocking effects, 
different quenchers are not allowed to occupy the same site. 
Therefore, in this case the number of the sites occupied by quenchers is $N$. 
In other words, 
the number of the sites occupied by quenchers is generally larger 
in the presence of site blocking effects than in its absence at any time. 
Accordingly, 
the probability that the sites occupied by quenchers will include the target site 
is higher in the presence of site blocking effects than in its absence at any time. 
As a result,  
the survival probability of the target decays faster in the presence of site blocking 
effects than in its absence. 

It is interesting to note the quite opposite effect of site blocking on   
the survival probability of a geminate pair with a large initial separation. 
Recently, it has been shown that 
the pair survival probability decays 
slower in the presence of site blocking effects by inert particles. \cite{Schmit} 
Here, 
the diffusion toward the target is just hindered 
by inert gases. 

Finally, we comment on the excluded volume interaction between the target and a quencher. 
It is possible to exclude the origin occupied by the target for the random walk of quenchers by
modifying the lattice Green's function of periodic lattice. 
The research in this direction is now undertaken.

%%%%%%%%%%%%%%%%%%%%%%%%%%%%%%%%%%%%%%%%%%%%%%%%%%%%%%%%%%%%%%%%%%%%%%
% ACKNOWLEDGMENTS %%%%%%%%%%%%%%%%%%%%%%%%%%%%%%%%%%%%%%%%%%%%%%%%%%%%
%%%%%%%%%%%%%%%%%%%%%%%%%%%%%%%%%%%%%%%%%%%%%%%%%%%%%%%%%%%%%%%%%%%%%%
%\newpage
%\noindent{\Large\bf Acknowledgment}
%\vspace{0.5cm}
\acknowledgments

We would like to thank Prof. K. Kitahara 
for many useful discussions. 

%%%%%%%%%%%%%%%%%%%%%%%%%%%%%%%%%%%%%%%%%%%%%%%%%%%%%%%%%%%%%%%%%%%%%%
% Appendixes %%%%%%%%%%%%%%%%%%%%%%%%%%%%%%%%%%%%%%%%%%%%%%%%%%%%%%%%%
%%%%%%%%%%%%%%%%%%%%%%%%%%%%%%%%%%%%%%%%%%%%%%%%%%%%%%%%%%%%%%%%%%%%%%
%\newpage
%\noindent{\Large\bf Acknowledgment}
%\vspace{0.5cm}
\appendix
\section{Proof of $1 > ck_0\hat{G}_0(\vec{r}_R,\vec{r}_R,z) >0$}
\label{appA}
We first prove 
$1/\hat{G}_0(\vec{r}_R,\vec{r}_R,z) -ck_0>0$ which 
can be  transformed into 
$1 > ck_0\hat{G}_0(\vec{r}_R,\vec{r}_R,z) $. 
By introducing 
\begin{align}
\hat{G}_0(\vec{r}_R,\vec{r}_R,z)=\frac{1- \hat{\psi} (z)}{z} U(z), 
\nonumber 
\end{align}
we obtain 
\begin{align}
1/\hat{G}_0(\vec{r}_R,\vec{r}_R,z) -ck_0 
&= \frac{ck_0 \left(1- U(z)\right)+s +1/\tau_0 +\Gamma}{U(z)} . 
\label{SVtcslocal1_2}
\end{align}
From the definition of $U(s)$ given by Eq. (\ref{latticeU}), 
we can show $U(z) \geq 1$ since 
the denominator in the integrand of $U(s)$,  
$1 - \hat{\psi} (z) \tilde{\lambda} (\vec{k})$, is smaller than $1$. 
Since the denominator of Eq. (\ref{SVtcslocal1_2}) is positive, 
we need to prove the positivity of the numerator,  
$ck_0 \left(1- U(z)\right) +\left(1/\tau_0 \right)+\Gamma >0$. 
By using  
\begin{align}
1- U(z) &= \frac{1}{(2\pi)^d} \int \cdots \int_{-\pi}^{\pi} d^d \vec{k}
\frac{- \Gamma \tilde{\lambda} (\vec{k})}{z+\Gamma \left[ 1-  \tilde{\lambda} (\vec{k}) \right]} ,
\label{SVtcslocal_ap1}
\end{align}
the numerator of Eq. (\ref{SVtcslocal1_2}) can be rewritten as, 
\begin{align}
\Gamma + ck_0 \left(1- U(z)\right) 
&= \frac{1}{(2\pi)^d} \int \cdots \int_{-\pi}^{\pi} d^d \vec{k}
\frac{\Gamma \left(s+1/\tau_0 \right) + 
\Gamma \left( c k_0 + \Gamma \right) 
\left( 1-\tilde{\lambda} (\vec{k})\right) 
}
{ s+1/\tau_0  +ck_0 
+\Gamma \left(1-\tilde{\lambda} (\vec{k})\right) } . 
\label{SVtcslocal_ap2}
\end{align}
Since  
$\tilde{\lambda} (\vec{k}) \equiv \frac{1}{2d} \sum_{j=1}^{2d} \cos \left(\vec{k} \cdot \vec{b}_j/b \right) < 1$, 
Eq. (\ref{SVtcslocal_ap2}) is positive.  
Therefore, $ck_0 \left(1- U(z)\right) +\left(1/\tau_0 \right)+\Gamma >0$ and 
it leads to
$
1/\hat{G}_0(\vec{r}_R,\vec{r}_R,z)-c k_0 >0
$, 
which can be rewritten as, 
\begin{align}
1> c k_0 \hat{G}_0(\vec{r}_R,\vec{r}_R,z) >0,  
\label{ap_con1}
\end{align}
where we have used the fact that 
both $ c k_0$  and $\hat{G}_0(\vec{r}_R,\vec{r}_R, z)$ 
are positive. 
$z$ in Eq. (\ref{ap_con1}) is given by 
$z=s+1/\tau_0 + ck_0$. 
Therefore, if we take the limit of $s\rightarrow 0$ in Eq. (\ref{ap_con1}) 
we have Eq. (\ref{conditionl1}).


\begin{references}

\bibitem{Rice}
S. A. Rice, 
{\it Diffusion-Limited Reactions}, in
{\it Comprehensive Chemical Kinetics}, edited by
Bamford C H, Tipper C F H, and Compton R G,  
Vol.~25 (Elsevier, Amsterdam, 1985)
and references cited therein.

\bibitem{Inokuti} M. Inokuti  and F. Hirayama, 
{\it J. Chem. Phys.}  {\bf 43}, 1978 (1965).
%Influence of Energy transfer by the exchange mechanism on donor luminescence 
% Inokuti, M.; Hirayama, F.; , 
%J. Chem. Phys. Lett.  {\bf 1965}, {\it 43}, 1978.

\bibitem{Tachiya74} M.Tachiya and A. Mozumder, 
{\it Chem. Phys. Lett.}  {\bf 28}, 87 (1974).

\bibitem{Tachiya83} M. Tachiya, 
{\it Radiat. Phys. Chem.} {\bf 21}, 167 (1983).
% Theory of diffusion-controlled reactions: Formulation of the bulk reaction rate in terms of the pair probability
%Tachiya, M.; 
%Radiat. Phys. Chem. {\bf 21}, 167 (1983).
%{\it Radiat. Phys. Chem.} {\bf 1983}, {\it 21}, 167. 

\bibitem{Szabo89} 
A. Szabo, {\it J. Phys. Chem.} {\bf 93}, 6929 (1989).

\bibitem{Nakazato} K. Nakazato and K. Kitahara,
{\it Prog. Theor. Phys.} {\bf 64}, 2261 (1980).
% Site bloching effect in tracer diffusion on a lattice
%Nakazato, K.; Kitahara, K.; 
%K. Nakazato and K. Kitahara,
%{\it Prog. Theor. Phys. } {\bf 1980}, {\it 64}, 2261. 

%%% revised 
\bibitem{Kuzovkov} V. N. Kuzovkov, E. A. Kotomin, 
{\it J. Chem. Phys.} {\bf 98}, 9107, (1993);  
V. N. Kuzovkov, E. A. Kotomin, and W. von Niessen, 
{\it J. Chem. Phys.} {\bf 105}, 
9486 (1996); 
V. N. Kuzovkov, E. A. Kotomin, and W. von Niessen, 
{\it Phys. Rev. E} {\bf 54}, 6128 
(1996). 
%%% revised 

\bibitem{Arfken}
G. B. Arfken and H. J. Weber, {\it Mathematical Methods for Physicists, 4th ed.} (Academic
Press, San Diego, 1995).

\bibitem{Allinger} K. Allinger and A. Blumen,
{\it J. Chem. Phys.} {\bf 72}, 4608 (1980).
% On the direct energy transfer to moving acceptors 
%Allinger, K.; Blumen, A.; 
%K. Allinger and M. Tachiya,
%
%{\it J. Chem. Phys. } {\bf 1980}, {\it 72}, 4608. 

\bibitem{Suzuki} Y. Suzuki, K. Kitahara, Y. Fujitani, and S. Kinouchi,
{\it J. Phys. Soc. Jpn.} {\bf 71}, 2936 (2002).
%Vacancy-Assisted Diffusion in a Honeycomb Lattice and in a Diamond Lattice
% Suzuki, Y; Kitahara, K.; Fujitani, Y.; Kinouchi, S.; 
%Y. Suzuki, K. Kitahara, Y. Fujitani, and S. Kinouchi: 
%{\it J. Phys. Soc. Jpn.} {\bf 2002}, {\it 71}, 2936. 

\bibitem{Okamoto07} R. Okamoto and Y. Fujitani, 
{\it J. Phys. Soc. Jpn.} {\bf 74}, 2510 (2005).
%Group-Theoretical Calculation of the Diffusion Coefficient via the Vacancy-Assisted 
% Okamoto, R;  Fujitani, Y.; 
%R. Okamoto and Y. Fujitani: 
%{\it J. Phys. Soc. Jpn.} {\bf 2005}, {\it 74}, 2510. 

\bibitem{Doi} M. Doi,  
{\it J. Phys. A} {\bf 9}, 1479 (1976). 

%\bibitem{Kotomin} E. Kotomin, and V. Kuzovkov, 
% (Elsevier, North Holland, Amsterdam, ...
%{\it Modern Aspects of Diffusion-Controlled Reactions: Cooperative Phenomena in Bimolecular Processes, Vol. 34 of Comprehensive Chemical Kinetics}, 
%((Elsevier, North Holland, Amsterdam, 1996).

\bibitem{Kotomin} V. Kuzovkov and E. Kotomin, {\it Rep. Prog. Phys.} {\bf 51}, 1479 (1988). 

\bibitem{Peliti} L. Peliti, {\it J. Physique} {\bf 46}, 1469 (1985). 

\bibitem{vanKampen}
N. G. Van Kampen, {\it Stochastic Processes in Physics and Chemistry, 2nd ed.} (North-Holland, Amsterdam, 1992). 

\bibitem{Benichou}
O. B\'{e}nichou and G. Oshanin, 
{\it Phys. Rev. E} {\bf }66, 031101 (2002)
%Ultraslow vacancy-mediated tracer diffusion in two dimensions: The Einstein relation verified

\bibitem{vanBeijeren85}
H. van Beijeren and R. Kutner, {\it Phys. Rev. Lett.} {\bf 55}, 238  (1985).
%Mean square displacement of a tracer particle in a hard-core lattice gas(logarithmic correction for square lattice 

\bibitem{SzaboPRL} A. Szabo, R. Zwanzig, and N. Agmon,  
{\it Phys. Rev. Lett.} , {\bf 61}, 2496 (1988). 
%mobile traps
%{\it Phys. Rev. Lett.} {\bf 1988}, {\it 61}, 2496. 

\bibitem{Bramson}
M. Bramson and J. L. Lebowitz, {\it Phys. Rev. Lett.} {\bf 61}, 2397
(1988); {\bf 62}, 694 (1989); {\it J. Stat. Phys.} {\bf 62}, 297 (1991).

\bibitem{Bhatia}
D. P. Bhatia, M. A. Prasad and D. Arora, {\it Phys. Rev. Lett. } {\bf 75}, 586 (1995).
%Comment on "Pair and Triple correlations in the A + B -> B Diffusion-Controlled Reaction

\bibitem{Arora}
D. Arora, D. P. Bhatia and M. A. Prasad, {\it J. Stat. Phys.} {\bf 84}, 697 (1996).
%No. 3/4 697-711Survival probability in one dimension for theA+B¢ò?B reaction with hard-core repulsion

\bibitem{Burlatsky}
S. F. Burlatsky, M. Moreau, G. Oshanin and A. Blumen, {\it  Phys. Rev. Lett. } {\bf 75}, 585 (1995).
%Comment on "Pair and Triple correlations in the A + B -> B Diffusion-Controlled Reaction

\bibitem{Sokolov} I.M. Sokolov, R. Metzler, K. Pant, and M. C. Williams, {\it Phys. Rev. E} {\bf 72}, 041102 (2005).
% First passage time of N excluded-volume particles on a line
%Sokolov, I.M.; Metzler, R.; Pant, K.; Williams, M.C.; 
%{\it Phys. Rev. E} {\bf 2005}, {\it 72}, 041102. 

\bibitem{Sano}
H. Sano and M. Tachiya, {\it J. Chem. Phys.} {\bf 71}, 1276  (1979).

\bibitem{Hughes}
B. D. Hughes,
{\it Random Walks and Random Environments} vol. 1 (Clarendon Press, Oxford, 1995)
and references cited therein.

\bibitem{Abramowitz} M. Abramowitz and I. A. Stegun, 
{\it Handbook of Mathematical Functions}, 
(Dover, New York, 1972).

\bibitem{Lee2000} 
J. Lee, J. Sung, and S. Lee, 
{\it J. Chem. Phys.} {\bf 113}, 8686 (2000).
%Excluded volume effects on the diffusion-influenced reaction: The many particle
%kernel approach
%VOLUME 113, NUMBER 19 15 NOVEMBER 2000 8686-8692
%{\it J. Chem Phys. } {\bf 2000}, {\it 113}, 8686. 

\bibitem{Shin2003} 
J. Park, H. Kim, and K.J. Shin, 
{\it J. Chem. Phys.} {\bf 118}, 9697 (2003). 
%Park J, Kim H, Shin KJ
%Excluded volume effect on diffusion-influenced reactions in one dimension 
%JOURNAL OF CHEMICAL PHYSICS 118 (21): 9697-9703 JUN 1 2003 
%{\it J. Chem Phys. } {\bf 2003}, {\it 118}, 9697. 

\bibitem{Zumofen_EV}
G. Zumofen, A. Blumen, and J. Klafter, {\it Chem. Phys. Lett.} {\bf 117}, 340 (1985). % excluded volume effects in annihilation reactions

\bibitem{Schmit}
J. D. Schmit, E. Kamber, and J. Kondev, {\it Phys. Rev. Lett.} {\bf 102}, 218302 (2009). 
% Lattice Model of Diffusion-Limited Bimolecular Chemical Reactions in Confined Environments

\end{references}
\end{document}